\documentclass[prb,twocolumn,amsmath,amssymb,showpacs,superscriptaddress]{revtex4-2}
\usepackage{graphicx}
\usepackage{color}
\usepackage{hyperref}
\usepackage{upgreek}
\usepackage{braket}

\hypersetup{colorlinks=true, linkcolor=blue, citecolor=blue, urlcolor=blue}
\begin{document}

\title{Third-order optical response in d-wave altermagnets: Analytical and numerical results from microscopic model}

\author{Shihao Zhang*} 
\affiliation{School of Physics and Electronics, Hunan University, Changsha 410082, China} 
\email{zhangshh@hnu.edu.cn}

\begin{abstract}
Altermagnets represent a novel category of magnetic materials characterized by zero net magnetization yet featuring spin-split band structures, and they demonstrate distinctive orbital-spin locking phenomena. Commencing from the minimal multi-orbital tight-binding Hamiltonian of d-wave altermagnets, we conduct an analysis of the general formulas for the third-order injection and shift currents. These currents are solely determined by the quantum metric and quantum connection, being free from Berry curvature contamination. In the ideal scenario where the $\delta$-bond hopping $V_\delta$ approaches zero ($V_\delta = 0$), we derive closed-form analytical solutions for the third-order photoconductivities. For the general situation with a finite value of $V_\delta$, we present a perturbative analytical solution within the limit of $V_\delta \ll V_\pi$, and this solution is verified through numerical calculations. Our research establishes a comprehensive theoretical description of the third-order optospintronic responses in d-wave altermagnets based on a microscopic model. Moreover, it offers a viable approach for the experimental observation of pure quantum geometric effects. 
\end{abstract}

\maketitle

\section{INTRODUCTION}
Quantum geometry, described by the quantum metric (real part) and the Berry curvature (imaginary part) of the quantum geometric tensor, has emerged as the fundamental framework for understanding nonlinear optical responses in condensed matter systems \cite{ahn2022riemannian,holder2020consequences}. Over the past decades, the physical manifestations of the Berry curvature have been comprehensively investigated across a wide range of topological materials and magnetic systems, yielding numerous breakthroughs in optoelectronic and spintronic device applications. In stark contrast, the direct experimental detection of intrinsic effects originating purely from the quantum metric has remained a formidable, long-standing challenge in the field\cite{kim2025direct}. This bottleneck arises primarily from the ubiquitous coexistence of quantum metric and Berry curvature contributions in the vast majority of known material systems, where their mutual interference inevitably obscures the pure quantum geometric signals in optical measurements.

Recently, altermagnets have attracted significant attention as a new paradigm of magnetic materials, which combine the zero net magnetization of antiferromagnets with the spin-split band structures of ferromagnets \cite{gRuO2, bHalleffect,aCe4Sb3,bCe4Sb3,xiao2023spin,jiang2023enumeration,chen2023spin,ren2023enumeration,gao2023ai,qu2024extremely,tan2024bipolarized,guo2024valley,he2023nonrelativistic,okugawa2018weakly,hayami2019momentum,PhysRevB.101.220403,PhysRevB.102.144441,PhysRevB.75.115103,zhang2025intrinsic,aMnF2,bMnF2,cMnF2,FeSb2,CrSb,CrSbexpriment,bCrSb,PhysRevLett.133.206401,liao2025direct,akCl,bkCl,Cr2SO,aRuF4,bRuF4,okugawa2018weakly,aV2Se2O,bV2Se2O,c2fq-hkk4,duan2025antiferroelectric,zhang2025sliding,wang2025spin}. The optical properties of altermagnets have attracted significant attention due to their potential applications in opto-spintronics \cite{vila2025orbital, hariki2024magneto}. In particular, Vila \textit{et al.} \cite{vila2025orbital,li2026quantum,wang2026valley} demonstrated a unique orbital-spin locking in d-wave altermagnets, which leads to spin-selective optical absorption: linearly polarized light along the $x$ direction selectively excites spin-down electrons, while light along the $y$ direction selectively excites spin-up electrons. This effect originates from the coupling between orbitals and spins through the sublattice degree of freedom, mediated by the crystal field and antiferromagnetic exchange interaction. Despite these advances in understanding the linear optical properties of altermagnets, the interplay between microscopic atomic hopping parameters, quantum geometry, and higher-order nonlinear optical responses in these systems remains largely uncharted territory. In particular, the potential of d-wave altermagnets to serve as an ideal platform for probing pure quantum geometric effects in nonlinear optics has yet to be systematically explored.

In this work, we systematically study the third-order optical response effects in d-wave altermagnets. We first introduce the minimal tight-binding Hamiltonian of d-wave altermagnets, and derive the core quantum geometric quantities, including the Berry connection, quantum metric, and quantum connection. We then derive the closed-form analytical solutions about third-order injection and shift currents for this system in the ideal limit of vanishing $\delta$-bond hopping $V_\delta=0$. For the general case with finite $V_\delta$, we present the perturbative analytical solution in the weak $V_\delta$ limit, which is proved by numerical calculations. Additionally, we characterize the spin polarization of the third-order photocurrent, revealing its exceptional robustness against finite $V_\delta$ and a dramatic enhancement compared to the linear optical response regime.

\section{THEORETICAL MODEL AND QUANTUM GEOMETRY}
\subsection{Minimal Hamiltonian of d-wave altermagnets}
\begin{figure*}[t]
    \centering
    \includegraphics[width=1.0\textwidth]{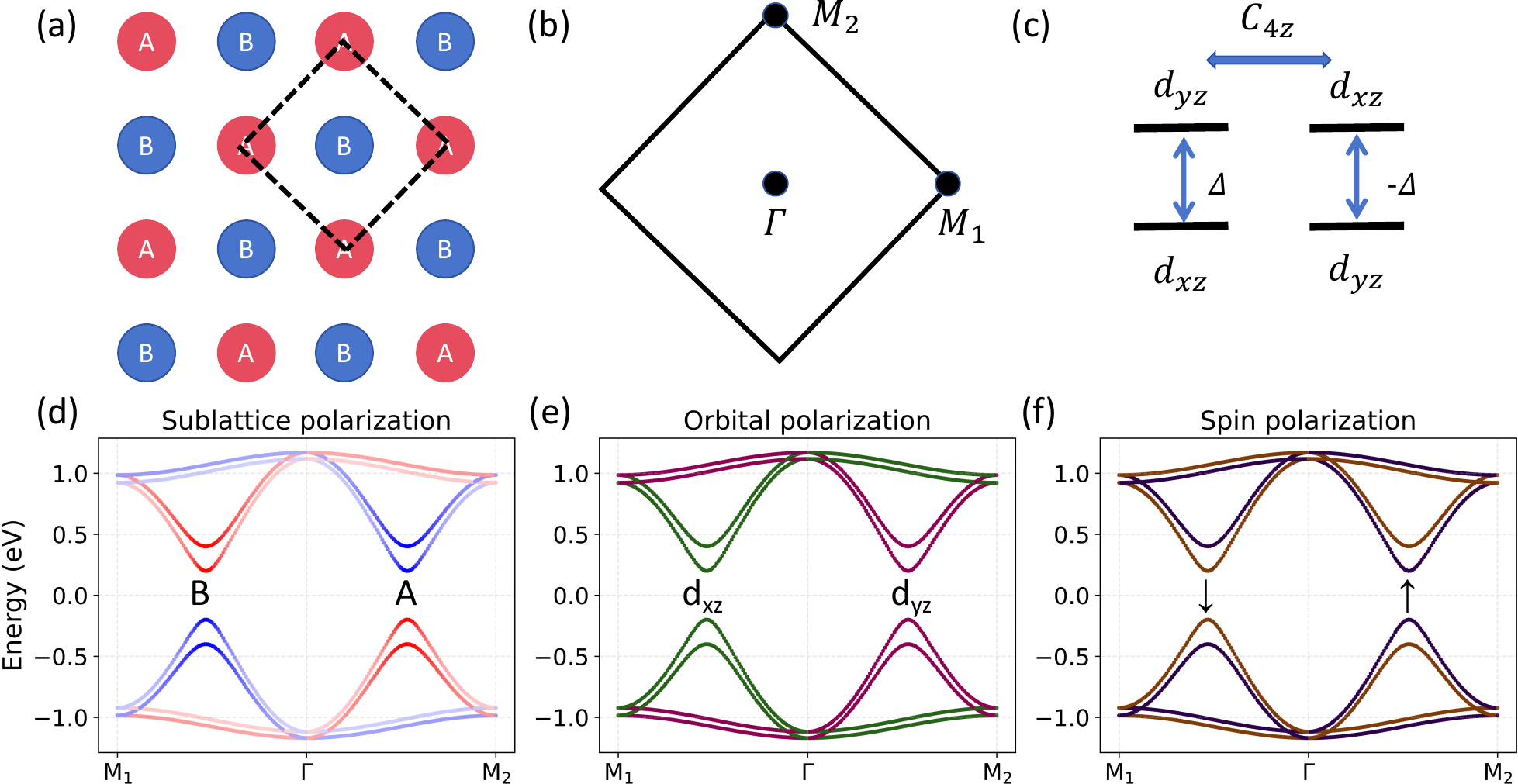}
    \caption{(a) The atomic structure of minimal model. The unitcell is remarked with black dashed lines. A and B sublattices carry opposite spins, and they are related with $C_{4z}$ symmetric operation. (b) The Brillouin zone of minimal model. Here $M_1=(\pi/a,0)$ and $M_2=(0,\pi/a)$. (c) Crystalline field induced splitting between $d_{xz}$ and $d_{yz}$ orbitals which are opposite in the different sublattices. (d-f) The sublattice, orbital, and spin polarization in the energy bands obtained from minimal model. Here $V_\pi=1.0$ eV, $V_\delta=0.1$ eV, $\Delta=0.3$ eV, $m=0.1$ eV.}
    \label{fig3}
\end{figure*}
We start with the minimal multi-orbital tight-binding Hamiltonian for d-wave altermagnets\cite{vila2025orbital}, 
\begin{equation}
\mathcal{H} = \mathcal{H}_0 + \mathcal{H}_{CF} + \mathcal{H}_{ex},
\end{equation}
where $\mathcal{H}_0$ describes the nearest-neighbor hopping, $\mathcal{H}_{CF}$ is the crystal field term, and $\mathcal{H}_{ex}$ is the antiferromagnetic exchange term. In the basis of sublattice ($\sigma$), orbital ($\tau$), and spin ($s$) degrees of freedom, these terms can be written as:
\begin{align}
\mathcal{H}_0 &= \sigma_x \otimes \left[ A(k)\tau_0 + B(k)\tau_z \right] \otimes s_0, \\
\mathcal{H}_{CF} &= \Delta \sigma_z \otimes \tau_z \otimes s_0, \\
\mathcal{H}_{ex} &= m \sigma_z \otimes \tau_0 \otimes s_z,
\end{align}
where
\begin{align}
A(k) &= \frac{V_\pi + V_\delta}{2} \left( \cos k_x a + \cos k_y a \right), \\
B(k) &= \frac{V_\pi - V_\delta}{2} \left( \cos k_x a - \cos k_y a \right).
\end{align}

Here, $V_\pi$ and $V_\delta$ are the Slater-Koster bond integrals for $\pi$ and $\delta$ bonds, respectively, $\Delta$ is the crystal field strength, $m$ is the exchange interaction strength, and $a$ is the distance between A and B sublattices. The Pauli matrices $\sigma$, $\tau$, and $s$ act on the sublattice, orbital, and spin spaces, respectively. Usually $V_\delta$ is much smaller than $V_\pi$. Here spin-orbit coupling (SOC) effect is always weak and thus not taken into consideration. For example, in the Fe$_2$Se$_2$O monolayer, the magnetic anisotropy energy induced by SOC effect is only 0.8\,meV\cite{fe2se2o}, but the magnetic exchange coupling reaches 0.2$\sim$1.1\,eV in the M$_2$Se$_2$O monolayer\cite{zhang2025sliding} (M is transition metal element).

The Hamiltonian commutes with $s_z$, so it can be block-diagonalized into two 4-dimensional subspaces corresponding to spin-up ($s=+1$) and spin-down ($s=-1$). Furthermore, there is no coupling between the $d_{xz}$ and $d_{yz}$ orbitals, so each 4-dimensional subspace can be further block-diagonalized into two independent 2-dimensional subspaces:
\begin{align}
\mathcal{H}_s^{xz}(k) &= C^{xz}(k)\sigma_x + D_s^{xz}\sigma_z, \\
\mathcal{H}_s^{yz}(k) &= C^{yz}(k)\sigma_x + D_s^{yz}\sigma_z,
\end{align}
where
\begin{align}
C^{xz}(k) &= V_\pi \cos k_x a + V_\delta \cos k_y a, \quad D_s^{xz} = \frac{ms}{2} + \Delta, \\
C^{yz}(k) &= V_\delta \cos k_x a + V_\pi \cos k_y a, \quad D_s^{yz} = \frac{ms}{2} - \Delta.
\end{align}

A crucial property of this Hamiltonian is that it is \textit{entirely real} for all $k$. This has profound implications for the quantum geometry of the system. The Hamiltonian also possesses $C_{4z}T$ symmetry, where $C_{4z}$ is a fourfold rotation about the $z$ axis and $T$ is time-reversal symmetry. This symmetry ensures that the electronic structure is invariant under the transformation $(k_x, k_y) \to (-k_y, k_x)$. As shown in the Fig.\,1, different sublattices contribute to different orbitals electrons with opposite spins.

Thus, in the following discussions, we handle the standard two-band form for each spin and $d_{xz}$ orbital channel :
\begin{align}
\mathcal{H}(\boldsymbol{k}) &= h_x(\boldsymbol{k}) \sigma_x + D_s \sigma_z,\\
h_x(\boldsymbol{k}) &= V_\pi \cos(k_x a) + V_\delta \cos(k_y a),\\
D_s &= \frac{ms}{2} + \Delta.
\label{eq:hamiltonian}
\end{align}
For the $d_{yz}$ orbital channel, the form of $h_x(\boldsymbol{k})$ is symmetric with $k_x \leftrightarrow k_y$, and $D_s$ is replaced by $D_s' = \frac{ms}{2} - \Delta$.

The eigenvalues of the Hamiltonian \eqref{eq:hamiltonian} are
\begin{equation}
E_{\pm}(\boldsymbol{k}) = \pm h(\boldsymbol{k}), \quad h(\boldsymbol{k}) = \sqrt{h_x^2(\boldsymbol{k}) + D_s^2},
\label{eq:eigenvalue}
\end{equation}
where $|+\rangle$ denotes the unoccupied conduction band and $|-\rangle$ denotes the fully occupied valence band at zero temperature. The interband transition frequency is
\begin{equation}
\omega_{+-}(\boldsymbol{k}) = \frac{E_+(\boldsymbol{k}) - E_-(\boldsymbol{k})}{\hbar} = \frac{2h(\boldsymbol{k})}{\hbar}.
\label{eq:omega}
\end{equation}
The system is a gapped insulator with the minimum band gap $E_{g,\text{min}} = 2|D_s|$, and the Fermi level $E_F=0$ lies in the middle of the band gap, leading to the Fermi distribution difference $f_{-+} = f(E_-) - f(E_+) = 1$ at zero temperature.

A crucial property of this Hamiltonian is that it is entirely real for all $\boldsymbol{k}$, which leads to identically zero Berry curvature over the entire BZ, eliminating all Berry curvature-mediated optical responses. This makes the system an ideal platform for studying pure quantum geometric effects.

We emphasize that the vanishing Berry curvature itself is not unique to d-wave altermagnets; it is a general property of all systems in which spin-orbit coupling can be neglected. What makes d-wave altermagnets truly unique is that they are the only known systems that combine vanishing Berry curvature with inversion symmetry and broken spin degeneracy. This combination is essential for observing pure quantum metric effects, as it eliminates both Berry curvature contamination and spin degeneracy cancellation, while suppressing all competing second-order responses.

\subsection{Quantum geometric quantities}
In this section, we discuss the quantum geometry about our two-band system. The Berry connection matrix element is defined as \cite{xiao2010berry,PhysRevB.104.085114}
\begin{equation}
r_{nm}^\mu = \bra{u_n} i\partial_{k_\mu} \ket{u_m},
\end{equation}
where $\ket{u_n}$ are the normalized eigenstates of the Hamiltonian \eqref{eq:hamiltonian}. For the real eigenstates of our system, the diagonal Berry connections are identically zero: $r_{++}^\mu = r_{--}^\mu = 0$, and the non-diagonal Berry connection is
\begin{equation}
r_{+-}^x = r_{-+}^x = i \frac{D_s \cdot \partial_{k_x} h_x}{2h^2(\boldsymbol{k})}.
\label{eq:berry}
\end{equation}

The quantum metric, which is the real part of the quantum geometric tensor, is given by
\begin{equation}
g_{\mu\nu}(\boldsymbol{k}) = r_{+-}^\mu \cdot r_{-+}^\nu.
\label{eq:metric}
\end{equation}
Substituting Eq. \eqref{eq:berry} into Eq. \eqref{eq:metric}, we obtain the quantum metric for the $d_{xz}$ orbital subsystem:
\begin{multline}
g^{xz,s}(k) \!= \!\frac{a^2 (D_s^{xz})^2}{4h_{xz}^4(k)}\\
\times \begin{pmatrix}
V_\pi^2 \sin^2 k_x a & V_\pi V_\delta \sin k_x a \sin k_y a \\
V_\pi V_\delta \sin k_x a \sin k_y a & V_\delta^2 \sin^2 k_y a
\end{pmatrix}.
\label{eq:metric_xz}
\end{multline}
Similarly, for the $d_{yz}$ orbital subsystem,
\begin{multline}
g^{yz,s}(k)\! =\! \frac{a^2 (D_s^{yz})^2}{4h_{yz}^4(k)}\\
\times\begin{pmatrix}
V_\delta^2 \sin^2 k_x a & V_\pi V_\delta \sin k_x a \sin k_y a \\
V_\pi V_\delta \sin k_x a \sin k_y a & V_\pi^2 \sin^2 k_y a
\end{pmatrix}.
\label{eq:metric_yz}
\end{multline}

Since the diagonal Berry connections are zero, the covariant derivative of the Berry connection reduces to the ordinary partial derivative:
\begin{equation}
\nabla_{k_x}^\ell r_{nm}^x = \frac{d^\ell}{dk_x^\ell} r_{nm}^x,
\end{equation}
where $\ell$ is the order of the derivative.

The quantum connection is defined as \cite{ezawa2025higher}
\begin{equation}
C_{xxx}^{mn} = r_{nm}^x \cdot \nabla_x r_{mn}^x,
\label{eq:qc1}
\end{equation}
and the $\ell$-th order higher-order quantum connection is
\begin{equation}
C_{xxxx}^{mn(\ell)} = \left( \nabla_x^\ell r_{mn}^x \right) \cdot \nabla_x r_{nm}^x.
\label{eq:qcl}
\end{equation}
These quantum connections are the core quantities that determine the higher-order shift currents in the system.

Now we discuss the nonlinear optical response in our system. Due to the inversion symmetry of the system, all even-order photoconductivities are strictly forbidden, and only odd-order responses are non-zero. In the $\ell$-th order optical response under applied electric field along $x$ direction, the photocurrent density along $c$ direction (c=x, y, z) is described as $j^c=\sigma^{x;x^\ell}E_x^\ell$. In this work, we focus on the $\ell$-order photocurrent from two ac electric fields and $\ell-2$ static electric fields.
\begin{equation}
    j^{x;x^{\ell}}=\sigma ^{x;x^{\ell}} E_x(\omega) E_x (-\omega) [E_x(0)]^{\ell-2}
\end{equation}
where $\omega$ is frequency of ac electric field. 

The injection current is derived from the equation of motion of Bloch electrons under external electric fields. In the microscopic viewpoint, the injection current originates from the difference between bands' difference. The general formula for the $\ell$-th order injection current photoconductivity is related to quantum metric \cite{ezawa2025higher,ezawa2026quantum,ezawa2025quantum}
\begin{equation}
\frac{\sigma_{\text{inject}}^{x;x^\ell}}{\sigma_{\text{inject}}^{(\ell)}} = \frac{1}{V} \sum_{\boldsymbol{k}} f_{-+} \frac{\partial^{\ell-1} \omega_{+-}}{\partial k_x^{\ell-1}} g_{xx} \delta(\omega_{+-} - \omega),
\label{eq:inject_general}
\end{equation}
where $V$ is the volume of the system, and the normalization constant is
\begin{equation}
\sigma_{\text{inject}}^{(\ell)} = 2\pi (\ell-1) \frac{e^{\ell+1}}{\hbar^\ell} \tau^{\ell-1},
\end{equation}
with $\tau$ being the electron relaxation time. The injection current is proportional to $\tau^{\ell-1}$, and thus dominates in the clean limit.

For a 2D system, the momentum summation is converted to the BZ integral:
\begin{equation}
\frac{1}{V} \sum_{\boldsymbol{k}} = \frac{1}{(2\pi)^2} \int_{\text{BZ}} d^2k.
\end{equation}
Substituting $\omega_{+-}=2h/\hbar$ and the delta function identity $\delta(ax+b)=\frac{1}{|a|}\delta(x+b/a)$ into Eq. \eqref{eq:inject_general}, we obtain the simplified integral form:
\begin{equation}
\sigma_{\text{inject}}^{x;x^\ell} = \frac{(\ell-1) e^{\ell+1} \tau^{\ell-1}}{2\pi \hbar^{\ell}} \int_{\text{BZ}} d^2k \frac{\partial^{\ell-1} h}{\partial k_x^{\ell-1}} g_{xx} \delta\left(h - \frac{\hbar\omega}{2}\right).
\label{eq:inject_integral}
\end{equation}

The shift current is induced when the mean positions are different between the different bands. The shift current can be derived from the perturbation expansion of the density matrix von Neumann equation. The general formula for the $\ell$-th order shift current photoconductivity is \cite{ezawa2025higher}
\begin{equation}
\sigma_{\text{shift}}^{x;x^\ell} = -\frac{e^2}{\hbar V} \left( \frac{e}{\hbar\omega} \right)^{\ell-2} \sum_{\boldsymbol{k}} f_{-+} r_{+-;x}^x \nabla_{k_x}^{\ell-2} r_{+-}^x \delta(\omega_{+-} - \omega),
\label{eq:shift_general}
\end{equation}
where $r_{+-;x}^x = \nabla_x r_{+-}^x$. The shift current is independent of the relaxation time $\tau$, and thus dominates in the dirty limit.

Substituting the quantum connection definition Eq. \eqref{eq:qcl} into Eq. \eqref{eq:shift_general}, we can rewrite the shift current in terms of the higher-order quantum connection:
\begin{equation}
\sigma_{\text{shift}}^{x;x^\ell} = -\frac{e^3}{\hbar^2 V} \left( \frac{e}{\hbar\omega} \right)^{\ell-2} \sum_{\boldsymbol{k}} f_{-+} C_{xxxx}^{+-(\ell-2)} \delta(\omega_{+-} - \omega).
\label{eq:shift_qc}
\end{equation}

\section{Results about third optical response under ideal limit $V_\delta=0$}
Because $V_\delta$ is much smaller than  $V_\pi$, we first consider the ideal limit $V_\delta=0$, where $h_x(\boldsymbol{k})=V_\pi\cos(k_x a)$ is only dependent on $k_x$, and the 2D integral can be reduced to a 1D integral via separation of variables:
\begin{equation}
\int_{\text{BZ}} d^2k = \frac{2\pi}{a} \int_{-\pi/a}^{\pi/a} dk_x.
\end{equation}

The resonance condition for optical transitions is $h(k_x)=\frac{\hbar\omega}{2}$, which gives
\begin{equation}
\cos^2(k_x a) = \frac{(\hbar\omega/2)^2 - D_s^2}{V_\pi^2} \equiv C(\omega).
\end{equation}
The response is non-zero only when $0\leq C(\omega)\leq1$, i.e., $2|D_s| \leq \hbar\omega \leq 2\sqrt{V_\pi^2+D_s^2}$, which is constrained by the Heaviside step function $\theta(\hbar\omega-2|D_s|)\theta(2\sqrt{V_\pi^2+D_s^2}-\hbar\omega)$.

\subsection{Third-order injection current ($\ell=3$, Jerk current)}
For $\ell=3$, substituting $\ell-1=2$ into Eq. \eqref{eq:inject_integral}, For $\ell=3$, the normalization constant is
\begin{equation}
\sigma_{\text{inject}}^{(3)} = 2\pi \cdot 2 \cdot \frac{e^4 \tau^2}{\hbar^3} = \frac{4\pi e^4 \tau^2}{\hbar^3}.
\end{equation}
The second-order derivative of the transition frequency is
\begin{equation}
\frac{\partial^2 \omega_{+-}}{\partial k_x^2} = \frac{2}{\hbar}\frac{\partial^2 h}{\partial k_x^2}.
\end{equation}
Then we obtain
\begin{align}
\sigma_{\text{inject}}^{x;x^3} &= \frac{4\pi e^4 \tau^2}{\hbar^3} \cdot \frac{1}{(2\pi)^2} \int_{\text{BZ}}d^2k \cdot \frac{2}{\hbar}\frac{\partial^2 h}{\partial k_x^2} g_{xx} \cdot \frac{\hbar}{2}\delta\left(h-\frac{\hbar\omega}{2}\right) \nonumber \\
&= \frac{e^4 \tau^2}{\pi \hbar^3} \int_{\text{BZ}} d^2k \, \frac{\partial^2 h}{\partial k_x^2} g_{xx} \delta\left(h-\frac{\hbar\omega}{2}\right).
\end{align}
Under $V_\delta=0$ limit, $h(k)$ depends only on $k_x$, so we get the one-dimensional core integral:
\begin{equation}
\sigma_{\text{inject}}^{x;x^3} = \frac{2 e^4 \tau^2}{a \hbar^3} \int_{-\pi/a}^{\pi/a} dk_x \, \frac{\partial^2 h}{\partial k_x^2} g_{xx} \delta\left(h-\frac{\hbar\omega}{2}\right).
\end{equation}

Let $\theta = k_x a$ (dimensionless momentum), so $h(\theta) = \sqrt{V_\pi^2 \cos^2\theta + D_s^2}$, $\frac{d}{dk_x} = a\frac{d}{d\theta}$. We note that
\begin{align}
\frac{dh}{dk_x} &= -\frac{a V_\pi^2 \sin\theta \cos\theta}{h},\\
\frac{d^2h}{dk_x^2} &= -\frac{a^2 V_\pi^2}{h^3} \cdot \left[ \cos2\theta \cdot h^2 + V_\pi^2 \sin^2\theta \cos^2\theta \right].
\end{align}

At resonance $h=\frac{\hbar\omega}{2}$, we have $\cos^2\theta_0 = \frac{(\hbar\omega/2)^2 - D_s^2}{V_\pi^2}$. The second-order derivative at resonance can be rewritten as
\begin{equation}
\frac{d^2h}{dk_x^2} = \frac{a^2}{h^3} \cdot \left[ D_s^2(V_\pi^2+D_s^2) - (\hbar\omega/2)^4 \right].
\end{equation}

There is a one-dimensional delta-function integral identity,
\begin{equation}
\int_{-\infty}^\infty dk \, f(k)\delta(g(k)-g_0) = \sum_{k_i: g(k_i)=g_0} \frac{f(k_i)}{\left|g'(k_i)\right|}.
\end{equation}

There are 4 symmetric resonance solutions $\theta_0$ in $\theta\in[-\pi,\pi]$, so the total contribution is 4 times the single-resonance contribution. And we have
\begin{equation}
\frac{|\sin\theta_0|}{|\cos\theta_0|} = \frac{\sqrt{V_\pi^2+D_s^2 - (\hbar\omega/2)^2}}{\sqrt{(\hbar\omega/2)^2 - D_s^2}}
\end{equation}
we obtain the closed-form analytical solution for the third-order injection current photoconductivity:
\begin{multline}
\sigma_{\text{inject}}^{x;x^3}(\omega) = \frac{128 e^4 a^2  D_s^2 \tau^2}{\hbar^9 \omega^6} \\
\times \frac{[(V_\pi^2+D_s^2)D_s^2-(\hbar\omega /2)^4]\sqrt{V_\pi^2 + D_s^2 - (\hbar\omega/2)^2}}{\sqrt{(\hbar\omega/2)^2 - D_s^2} }.
\label{eq:3rd_inject}
\end{multline}

Thus, we have the $\sigma_{\text{inject,xz},\downarrow}^{x;x^3}(\omega)=\sigma_{\text{inject,yz},\uparrow}^{y;y^3}(\omega)=\sigma_{\text{inject}}^{x;x^3}(\omega)$. And $\sigma_{\text{inject,xz},\downarrow}^{y;y^3}(\omega)=\sigma_{\text{inject,yz},\uparrow}^{x;x^3}(\omega)=0$ because corresponding quantum metric is zero under $V_{\delta}=0$ limit. Thus, in the $V_{\delta}=0$ limit, $x$-direction electric field induce the third-order injection current (Jerk current) only from spin-down channel of $d_{xz}$ orbitals, and $y$-direction electric field induce the Jerk current only from spin-up channel of $d_{yz}$ orbitals. This photocurrent is 100\% spin-polarized, which reveals that the direction of incident light can tune the spin polarization of photocurrent.

Near the band edge $\hbar\omega \approx 2|D_s|$, we define $\hbar\Delta\omega = \hbar\omega - 2|D_s| \ll |D_s|$, and the solution simplifies to
\begin{equation}
\sigma_{\text{inject}}^{x;x^3} \approx \frac{2 e^4 a^2 \tau^2 V_{\pi}^2}{\hbar^3 D_s^2}\sqrt{\frac{V_{\pi}^2}{|D_s|\hbar\Delta\omega}-1},
\end{equation}
which exhibits a square-root singularity at the band edge, corresponding to the van Hove singularity in the density of states. From the approximated solution, we also note that $\sigma_{\text{inject}}^{x;x^3} \propto V_{\pi}^3$ near the band edge.


\subsection{Third-order shift current ($\ell=3$)}
For the third-order shift current, We first compute the momentum derivative of the Berry connection $\partial_{k_x} r_{+-}^x$:
\begin{equation}
\partial_{k_x} r_{+-}^x = a \partial_\theta r_{+-}^x = -i \frac{a^2 V_\pi D_s}{2} \partial_\theta \left( \frac{\sin\theta}{h^2} \right).
\end{equation}
Because
\begin{align}
\partial_\theta \left( \frac{\sin\theta}{h^2} \right) = \frac{\cos\theta \left( h^2 + 2 V_\pi^2 \sin^2\theta \right)}{h^4},
\end{align}
we obtain
\begin{equation}
\partial_{k_x} r_{+-}^x = -i \frac{a^2 V_\pi D_s \cos\theta \left( h^2 + 2 V_\pi^2 \sin^2\theta \right)}{2 h^4}.
\end{equation}
Then we compute the first-order quantum connection $C_{xxx}^{+-}$
\begin{align}
C_{xxx}^{+-(1)} &= (\partial_{k_x} r_{+-}^x)^2 \nonumber \\
&= -\frac{a^4 V_\pi^2 D_s^2 \cos ^2\theta \left( h^2 + 2 V_\pi^2 \sin^2\theta \right)^2}{4 h^8}.
\label{eq:Cxxx}
\end{align}

\begin{figure*}[t]
    \centering
    \includegraphics[width=1.0\textwidth]{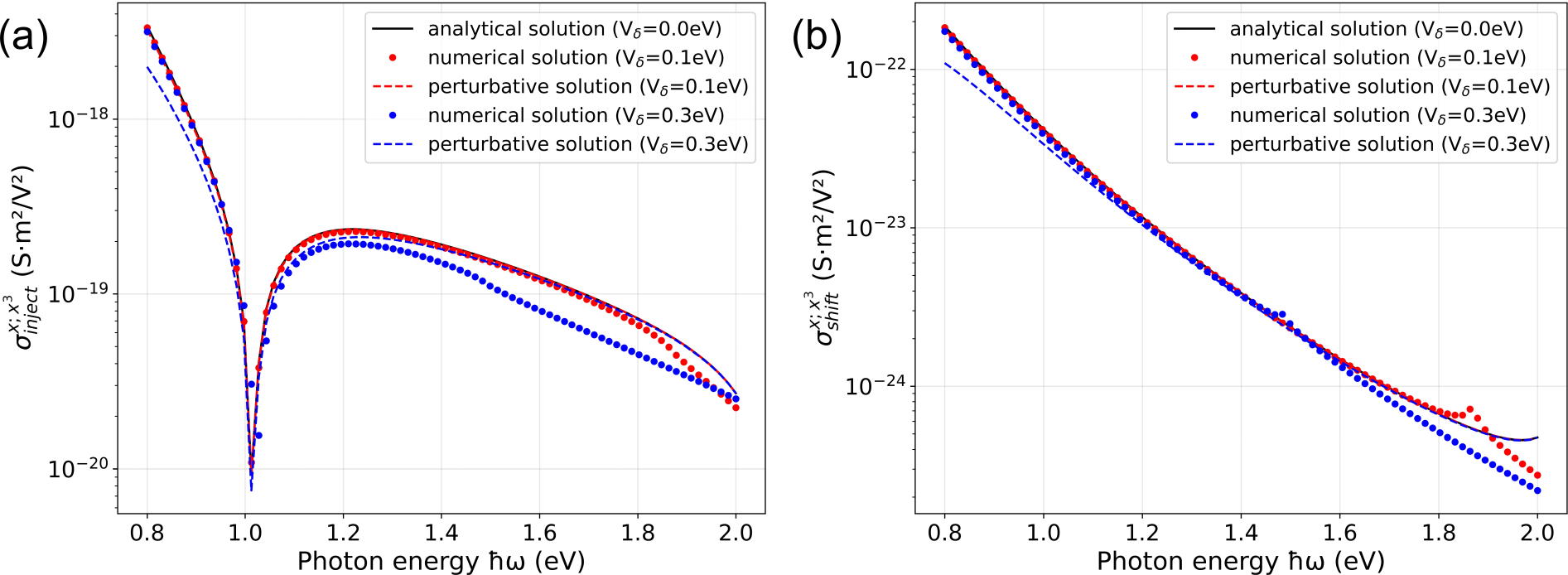}
    \caption{The analytical, numerical and perturbative solutions about third-order injection current (a) and shift current (b) conductivity. Here $V_\pi=1.0$ eV, $D_s=0.25$ eV, $a=0.4$ nm, $\tau=100$ fs. We consider two typical values of $V_\delta$: $0.1$ eV and $0.3$ eV, corresponding to $\varepsilon=0.1$ and $\varepsilon=0.3$, respectively. The analytical solutions are obtained from Eq.(\ref{eq:3rd_inject}) and Eq.(\ref{eq:corrected_33}). The numerical solutions are obtained from Eq.(\ref{eq:3rd_inject_line}) and Eq.(\ref{eq:correct_shift3_main}). And the perturbative solutions are obtained from Eq.(\ref{eq:perturb}).}
    \label{fig2}
\end{figure*}

In the ideal limit $V_\delta=0$, $h(k)$ depends only on $k_x$, so the integral becomes
\begin{align}
\sigma_{\text{shift}}^{x;x^3} = -\frac{e^4}{2\pi a \hbar^3 \omega} \int_{-\pi/a}^{\pi/a} dk_x \, C_{xxx}^{+-(1)} \, \delta(\omega_{+-} - \omega).
\label{eq:1D_integral}
\end{align}
Using the delta function scaling identity $\delta(ax + b) = \frac{1}{|a|}\delta(x + b/a)$, we rewrite the energy conservation delta function:
\begin{equation}
\delta(\omega_{+-} - \omega) = \delta\left( \frac{2h}{\hbar} - \omega \right) = \frac{\hbar}{2} \delta(h - h_0),
\end{equation}
where we define the resonance energy $h_0 = \frac{\hbar\omega}{2}$. Substituting into Eq. \eqref{eq:1D_integral}:
\begin{equation}
\sigma_{\text{shift}}^{x;x^3} = -\frac{e^4}{4\pi a \hbar^2 \omega} \int_{-\pi/a}^{\pi/a} dk_x \, C_{xxx}^{+-(1)} \, \delta(h - h_0).
\label{eq:delta_integral}
\end{equation}
Now we get
\begin{equation}
\sigma_{\text{shift}}^{x;x^3} = -\frac{4 e^4}{4\pi a \hbar^2 \omega} \frac{ C_{xxx}^{+-(1)}}{|dh/dk_x|} \bigg|_{h=h_0} = -\frac{e^4}{\pi a \hbar^2 \omega} \cdot \frac{C_{xxx}^{+-(1)}}{|dh/dk_x|} \bigg|_{h=h_0}.
\label{eq:resonance_sum}
\end{equation}
At resonance $h=h_0$, we use the identity $h_0^2 = V_\pi^2 \cos^2\theta + D_s^2$ to eliminate the $\theta$-dependence of the quantum connection. First, we simplify the term in Eq. \eqref{eq:Cxxx}:
\begin{align}
h^2 + 2 V_\pi^2 \sin^2\theta \bigg|_{h=h_0} &= h_0^2 + 2 V_\pi^2 (1 - \cos^2\theta) \nonumber \\
&= 2(V_\pi^2 + D_s^2) - h_0^2.
\label{eq:resonance_simplify}
\end{align}
This critical simplification eliminates all trigonometric dependence from the numerator of $C_{xxx}^{+-}$.

Next, we compute the derivative $\partial_{k_x} C_{xxx}^{+-}$ at resonance, and take the ratio with $|dh/dk_x|$. The trigonometric terms $\sin\theta \cos\theta$ from $C_{xxx}^{+-(1)}$ and $|dh/dk_x|$ cancel exactly, leaving:
\begin{equation}
\frac{C_{xxx}^{+-(1)}}{|dh/dk_x|} \bigg|_{h=h_0} = \frac{a^3 D_s^2 \left[ 2(V_\pi^2 + D_s^2) - h_0^2 \right]^2\sqrt{(\hbar\omega/2)^2 - D_s^2}}{4h_0^7\sqrt{V_\pi^2+D_s^2 - (\hbar\omega/2)^2}}.
\end{equation}
Then we obtain the final closed-form analytical solution for the third-order shift current:
\begin{multline}
\sigma_{\text{shift}}^{x;x^3}(\omega) = \frac{32e^4a^2 D_s^2}{\pi \hbar^9 \omega ^8} \\
\times \frac{\left[ 2(V_\pi^2 + D_s^2) - (\hbar \omega / 2)^2 \right]^2\sqrt{(\hbar\omega/2)^2 - D_s^2}}{\sqrt{V_\pi^2+D_s^2 - (\hbar\omega/2)^2}}.
\label{eq:corrected_33}
\end{multline}
Thus, the $\sigma_{\text{shift,xz},\downarrow}^{x;x^3}(\omega)=\sigma_{\text{shift,yz},\uparrow}^{y;y^3}(\omega)=\sigma_{\text{shift}}^{x;x^3}(\omega)$. And $\sigma_{\text{shift,xz},\downarrow}^{y;y^3}(\omega)=\sigma_{\text{shift,yz},\uparrow}^{x;x^3}(\omega)=0$ because corresponding quantum connection is zero under $V_{\delta}=0$ limit. As in the jerk-current cases, in the $V_{\delta}=0$ limit, an electric field along the $x$ direction drives a third-order shift current exclusively through the spin-down channel of the $d_{xz}$ orbitals., and $y$-direction electric field induce the shift current only from spin-up channel of $d_{yz}$ orbitals.

\section{Results about third optical response under general case: finite $V_\delta$}
For finite $V_\delta$, $h(\boldsymbol{k})$ is a function of both $k_x$ and $k_y$, and the resonance condition $h(\boldsymbol{k})=\frac{\hbar\omega}{2}$ corresponds to closed iso-frequency contours in the BZ. We use the 2D delta function identity to convert the area integral into a line integral over the iso-frequency contour $C(\omega)$:
\begin{equation}
\int\int_{\text{BZ}} d^2k F(\boldsymbol{k}) \delta\left(h(\boldsymbol{k})-\frac{\hbar\omega}{2}\right) = \oint_{C(\omega)} \frac{F(\boldsymbol{k})}{|\nabla_{\boldsymbol{k}}h(\boldsymbol{k})|} dl,
\label{eq:line_int}
\end{equation}
where $dl$ is the line element of the iso-frequency contour, and $|\nabla_{\boldsymbol{k}}h(\boldsymbol{k})|$ is the magnitude of the gradient of $h(\boldsymbol{k})$, given by
\begin{equation}
|\nabla_{\boldsymbol{k}}h(\boldsymbol{k})| = \frac{a |h_x|}{h} \sqrt{V_\pi^2 \sin^2(k_x a) + V_\delta^2 \sin^2(k_y a)}.
\end{equation}

\subsection{Exact line-integral form}
Using Eq. \eqref{eq:line_int}, we obtain the exact line-integral form of the third-order injection current for finite $V_\delta$:
\begin{equation}
\sigma_{\text{inject}}^{x;x^3}(\omega) = \frac{e^4 \tau^2 D_s^2 a^2 V_\pi^2}{4\pi \hbar^3} \oint_{C(\omega)} \frac{ \left( \frac{\partial^2 h}{\partial k_x^2} \right) \sin^2(k_x a) }{ h^4(\boldsymbol{k}) |\nabla_{\boldsymbol{k}}h(\boldsymbol{k})| } dl,
\label{eq:3rd_inject_line}
\end{equation}
where the iso-frequency contour $C(\omega)$ satisfies
\begin{equation}
V_\pi\cos(k_x a) + V_\delta\cos(k_y a) = \pm\sqrt{(\hbar\omega/2)^2 - D_s^2}.
\end{equation}
The contributions from the positive and negative branches are identical, so we can calculate only the positive branch and multiply by 2.

Similarly, the exact line-integral form of the third-order shift current is
\begin{multline}
\sigma_{\text{shift}}^{x;x^3}(\omega)= \frac{e^3 a^4 D_s^2 V_\pi^2}{32 \pi^2 \hbar \omega}\\
\times \oint_{C(\omega)}
\frac{\left[ 2 V_\pi h_x \sin^2(k_x a) + h^2(k) \cos(k_x a) \right]^2}
{h^8(k) \left|\nabla_{\boldsymbol{k}} h(k)\right|} dl.
\label{eq:correct_shift3_main}
\end{multline}
Because these line-integrals are elliptic integrals, we make numerical calculations about these integrals, and the results are shown in the Fig.\,2. The third-order shift current is numerically found to be about 4 orders of magnitude smaller than the injection current, which is consistent with the theoretical prediction that the injection current dominates in the clean limit.

\subsection{Perturbative analytical solution for $V_\delta \ll V_\pi$}
For realistic material parameters, $V_\delta \ll V_\pi$, we can perform a perturbative expansion with the small parameter $\varepsilon=V_\delta/V_\pi$. Taking the third-order injection current as an example, we expand the photoconductivity up to $\mathcal{O}(\varepsilon^2)$:
\begin{multline}
\frac{\sigma_{\text{inject}}^{x;x^3}(\omega)} {\sigma_{\text{inject},0}^{x;x^3}(\omega)}= \frac{\sigma_{\text{shift}}^{x;x^3}(\omega)} {\sigma_{\text{shift},0}^{x;x^3}(\omega)}= \\
\left[ 1 - \frac{\varepsilon^2}{2} \cdot \frac{V_\pi^2 + D_s^2 - (\hbar\omega/2)^2}{(\hbar\omega/2)^2 - D_s^2} + \mathcal{O}(\varepsilon^4) \right],
\label{eq:perturb}
\end{multline}
where $\sigma_{\text{inject},0}^{x;x^3}(\omega)$ and $\sigma_{\text{shift},0}^{x;x^3}(\omega)$ are the analytical solution for $V_\delta=0$ given in Eq. \eqref{eq:3rd_inject} and Eq. \eqref{eq:corrected_33}, respectively. As shown in the Fig.\,2, our perturbative solutions are well consistent with numerical solutions.

When we consider the spin-orbit coupling effect, the perturbative analytical solution can be written as
\begin{widetext}
\begin{equation}
        \frac{\sigma(\omega, \varepsilon, \lambda)}{\sigma_{00}(\omega)} = 1 - \frac{1}{2} \left( \frac{V_\delta}{V_\pi} \right)^2 \cdot \frac{V_\pi^2 + D_s^2 - (\hbar\omega/2)^2}{(\hbar\omega/2)^2 - D_s^2} - \frac{1}{2} \left( \frac{\lambda}{D_s} \right)^2 \cdot \frac{V_\pi^2}{(\hbar\omega/2)^2 - D_s^2} + \mathcal{O}(\varepsilon^4, \lambda^4)
    \label{eq:final_perturbative_solution}
\end{equation}
\end{widetext}
where $\lambda$ is spin-orbit coupling strength, and $\sigma_{00}(\omega)$ is the closed-form analytical solution in the ideal limit $V_\delta=0, \lambda=0$. From this result, we note that the influence of SOC strength can be neglected in the nonlinear optical response of wide-gap insulator.

\subsection{Spin polarization}
\begin{figure}[t]
    \centering
    \includegraphics[width=0.5\textwidth]{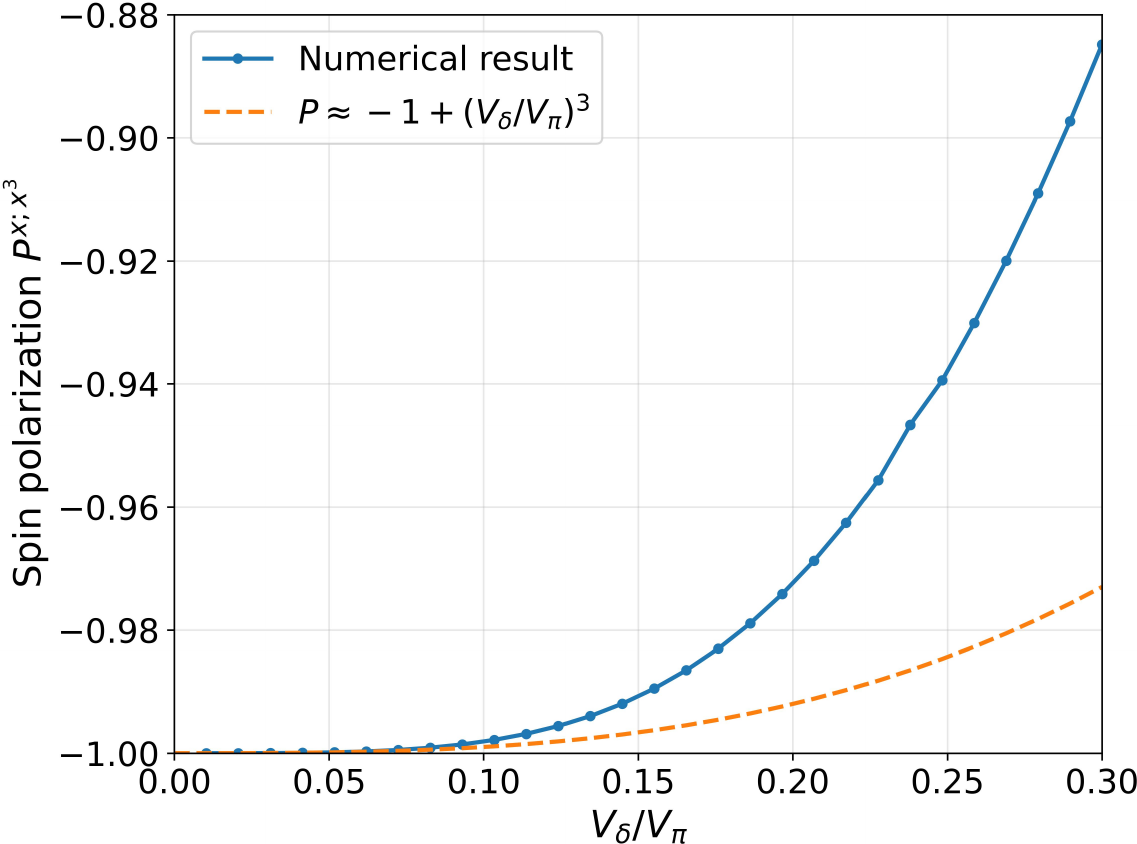}
    \caption{The numerical results about spin polarization in the Jerk current $P^{x;x^3}$. Here $V_\pi=1.0$ eV, $D_s=0.25$ eV, $a=0.4$ nm, $\tau=100$ fs.}
    \label{fig3}
\end{figure}
Vila \textit{et al.} \cite{vila2025orbital} uncovered a orbital–spin locking in $d$-wave altermagnets that drives spin-selective optical absorption: linearly polarized light along $x$ excites almost exclusively spin-down electrons, while polarization along $y$ picks out spin-up electrons. This spin filtering extends into the nonlinear regime. In the third-order response, the same selectivity emerges, but the injection current overwhelmingly dominates the shift current. We therefore concentrate on the third-order injection current as the primary channel for generating and analyzing spin polarization.
\begin{widetext}
\begin{align}
\sigma_{inject,xz\downarrow}^{x;x^3}(\omega) 
&= \frac{e^4 \tau^2 a^3 V_\pi^3 D_{xz\downarrow}^2}{4 \pi \hbar^3 h_0^6 C(\omega)} 
\oint_{C_{xz}(\omega)} \frac{\sin^2(k_x a) \left[ h_0^2 \left( V_\pi \sin^2(k_x a) - h_x^{xz}(k) \cos(k_x a) \right) - V_\pi \left(h_x^{xz}(k)\right)^2 \sin^2(k_x a) \right]}{\sqrt{V_\pi^2 \sin^2(k_x a) + V_\delta^2 \sin^2(k_y a)}} dl ,\\
\sigma_{inject,yz\uparrow}^{x;x^3}(\omega) 
&= \frac{e^4 \tau^2 a^3 V_\delta^3 D_{yz\uparrow}^2}{4 \pi \hbar^3 h_0^6 C(\omega)} 
\oint_{C_{yz}(\omega)} \frac{\sin^2(k_x a) \left[ h_0^2 \left( V_\delta \sin^2(k_x a) - h_x^{yz}(k) \cos(k_x a) \right) - V_\delta \left(h_x^{yz}(k)\right)^2 \sin^2(k_x a) \right]}{\sqrt{V_\delta^2 \sin^2(k_x a) + V_\pi^2 \sin^2(k_y a)}} dl .
\label{xzyzsigma}
\end{align}
\end{widetext}
Then we can conclude that the spin polarization $P$ of photocurrent from third-order optical response under small $V_{\delta}/V_{\pi}$ values can be written as
\begin{align}
    P^{x;x^3}&=\frac{j_{\uparrow}-j_{\downarrow}}{j_{\uparrow}+j_{\downarrow}}=\frac{\sigma ^{x;x^3} _{yz,\uparrow}-\sigma ^{x;x^3} _{xz,\downarrow}}{\sigma ^{x;x^3} _{yz,\uparrow}+\sigma ^{x;x^3} _{xz,\downarrow}} \sim -1+(\frac{V_{\delta}}{V_{\pi}})^3,\\
    P^{y;y^3}&=\frac{j_{\uparrow}-j_{\downarrow}}{j_{\uparrow}+j_{\downarrow}}=\frac{\sigma ^{y;y^3} _{yz,\uparrow}-\sigma ^{y;y^3} _{xz,\downarrow}}{\sigma ^{y;y^3} _{yz,\uparrow}+\sigma ^{y;y^3} _{xz,\downarrow}} \sim 1-(\frac{V_{\delta}}{V_{\pi}})^3.
    \label{approsp}
\end{align}
We further carry out numerical simulations of spin polarization based on Eq.\,(\ref{xzyzsigma}), with the results summarized in Fig.\,3. The approximation in Eq.\,(\ref{approsp}) holds quantitatively as long as $V_{\delta}/V_{\pi} < 5\%$. Remarkably, as shown in Fig.\,3, the photocurrent spin polarization remains above 88\% even for the relatively large ratio $V_\delta/V_\pi = 0.3$. Since the spin polarization in the first-order optical response scales as $[1 - (V_{\delta}/V_{\pi})]$, the spin polarization of the third-order photocurrent is dramatically enhanced compared with its first-order counterpart.

\section{CONCLUSION}
In this study, we systematically examine the third-order bulk photovoltaic effects in d-wave altermagnets, adhering to the standard theoretical framework for unconventional magnetic systems. For the ideal limit $V_\delta = 0$, we acquire closed-form analytical solutions for the third-order photoconductivities. In the general case where $V_\delta$ is finite, we present the exact line-integral form of the photoconductivities and derive the perturbative analytical solution in the weak $V_\delta$ limit. Our numerical findings demonstrate that the perturbative analytical solutions are in good agreement with the numerical solutions. Moreover, the photocurrent spin polarization exceeds 88\% even when $V_\delta/V_\pi = 0.3$, notably outperforming the first-order optical response. This remarkable characteristic offers a novel and efficient approach for all-optical spin injection in advanced optospintronic devices. 

Overall, our findings complete the theoretical framework of nonlinear optospintronics in altermagnets, and establish a full microscopic theoretical description of the third-order optical responses in d-wave altermagnets. This work not only provides a solid theoretical foundation for the experimental observation of pure quantum geometric effects, but also opens a new avenue for the design of novel altermagnet-based optoelectronic and spintronic devices.

\begin{acknowledgments}
This work was supported by the National Natural Science Foundation of China (No. 12304217), the National Key Research and Development Program of China (No. 2024YFA1410300), the Natural Science Foundation of Hunan Province (No. 2025JJ60002) and the Fundamental Research Funds for the Central Universities from China (No. 531119200247).
\end{acknowledgments}

\bibliographystyle{apsrev4-2}
\bibliography{reference}

\end{document}